\documentstyle[preprint,pra,aps]{revtex}
\begin{document}
\draft \title{\bf Relativistic effects in Ni~II and the search for
variation of the fine structure constant.}
\author{V. A. Dzuba, V. V. Flambaum, M. T. Murphy and J. K. Webb}
\address{School of Physics, University of New South Wales, UNSW Sydney NSW
2052, Australia}
\date{\today}
\maketitle

\begin{abstract}
Theories unifying gravity and other interactions suggest the possibility of
spatial and temporal variation of physical ``constants'' in the Universe.
Detection of high redshift absorption systems intersecting the sight lines
towards distant quasars provide a powerful tool for measuring these
variations. In the present paper we demonstrate that high sensitivity to
variation of the fine structure constant $\alpha$ can be obtained by
comparing cosmic and laboratory spectra of the Ni~II ion.  Relativistic
effects in Ni~II reveal many interesting features.  The Ni~II spectrum
exhibits avoided level crossing phenomenon under variation of $\alpha$ and
the intervals between the levels have strong nonlinear dependencies on
relativistic corrections.  The values of the transition frequency shifts,
due to the change of $\alpha$, vary significantly from state to state
including change of the sign.  This enhances the sensitivity to the
variation of $\alpha$ and reduces possible systematic errors.  The
calculations of $\alpha$-dependence of the nickel ion spectral lines that
are detectable in quasar absorption spectra have been performed using a
relativistic configuration interaction method.
\end{abstract} 
\vspace{1cm}
\pacs{PACS: 06.20.Jr , 31.30.Jv , 95.30.Dr}


\section{Introduction}

Possible variations of the fundamental physical constants in the expanding
Universe are currently of particular interest because of the implications
of unified theories, such as string theory and M-theory. They predict that
additional compact dimensions of space exist.  
The ``constants'' seen in our three-dimensional subspace of the theory will
vary according to any variation in the scale lengths of the extra compact
dimensions(see, e.g. \cite{Marciano84,Barrow87,Damour94}).
Gas clouds which intersect the sight lines towards distant quasars produce
absorption lines in astronomical spectra.  These absorption systems present
ideal laboratories in which to search for any temporal or spatial variation
of fundamental constants by comparing the observed atomic spectra from the
distant objects with laboratory spectra (see, e.g. \cite{Potechin95} and
references therein).

The energy scale of atomic spectra is given by the atomic unit
$\frac{me^4}{\hbar^2}$. In the non-relativistic limit, all atomic spectra
are proportional to this constant and analyses of quasar spectra cannot
detect any change of the fundamental constants.  Indeed, any change in the
atomic unit will be absorbed in the determination of the red shift
parameter $z$ ($1 + z = \frac{\omega}{\omega '}$, $\omega '$ is the
red-shifted frequency of the atomic transition and $\omega$ is the
laboratory value).  However, any change of the fundamental constants can be
found by measuring the relative size of relativistic corrections, which are
proportional to $\alpha^2$, where $\alpha = e^2/\hbar c$ is the fine
structure constant \cite{other}.
 
In our previous works \cite{Dzuba98,Dzuba99} we have demonstrated that high
sensitivity to the change of $\alpha$ can be achieved by comparing
transition frequencies of heavy and light atoms.  The results of our
calculations for Fe~II and Mg~II have been used in Ref.  \cite{Webb} where
the results of the search for $\alpha$-variation have been presented.
Applied to a sample of 30 absorption systems, spanning red-shifts
$0.5<z<1.6$, obtained with the Keck I telescope, the limits on variations
in $\alpha$ over a wide range of epochs have been derived.  For the whole
sample $\Delta \alpha /\alpha =-1.1\pm 0.4\times 10^{-5}$.  Whilst these
results are consistent with a time-varying $\alpha$, further work is
required to explore possible systematic errors in the data, although
careful searches have so far not revealed any \cite{Murphy1}. 
The obvious way to test
these results and further improve sensitivity is to include new atoms and
spectral lines with different frequencies and different dependence on
$\alpha$. It would be especially attractive to have lines with large
relativistic shifts of the opposite signs since the opposite signs of the
shifts lead to the suppression of the most dangerous systematic errors.
The shift of lines produced by systematic errors ``does not know''
about the signs of the relativistic shifts. Therefore, it is easier to 
eliminate systematic errors when the signs are different.

In the present paper we demonstrate that the Ni~II ion has a very
interesting spectrum which possesses these desirable properties (see Table
\ref{nickel}). It is also very important that there are several strong
Ni~II lines observed in the quasar absorption spectra.

Note that we present all results in this paper assuming that the atomic
unit of energy $\frac{me^4}{\hbar^2}$ is constant (since any variation of
this unit will be absorbed in the determination of the redshift parameter
$z$).

\section{Theory and results} 

Relativistic energy shift for a particular valence electron can be 
approximately described by the equation \cite{Dzuba99}
\begin{eqnarray}
	\Delta_n = \frac{E_n}{\nu}(Z\alpha)^2 \left[\frac{1}{j+1/2} -
	C(Z,j,l)\right],
\label{rel5}
\end{eqnarray}
where $\nu$ is the principal quantum number($E_n = -1/2 \nu^2$) and
$C(Z,j,l)$ accounts for the many-body effects. In many cases
$C(Z,j,l) \simeq 0.6$, however the accurate value of $C(Z,j,l)$ can
only be obtained from the many-body calculations. Formula (\ref{rel5})
accounts for the relativistic effects which are included into the
single-electron Dirac equation. Note that they cannot be reduced to
the spin-orbit interaction. For example, as is evident from the
formula (\ref{rel5}) the energy shift is large for $s$-electrons 
which have no spin-orbit interaction at all. Moreover, the spin-orbit
interaction doesn't even dominate in the relativistic energy shift. 
However, only the spin-orbit interaction can be found from the
analysis of the experimental fine structure splitting while
other relativistic effects remain ``hidden''. Note that the
Coulomb integrals which determine splitting between different
multiplets in many-electron states also contain relativistic corrections.

Thus, the analysis of the experimental atomic spectra does not
provide sufficient information about relativistic effects in
transition frequencies in atoms. For an atom with one external
electron above closed shells one can obtain an approximate relativistic
frequency shift by applying formula (\ref{rel5}) to both upper and lower 
states of the transition. For a many-electron atom like Ni~II this
procedure is too inaccurate. Therefore the only way to get the
results is to perform {\it ab initio} relativistic calculations.
However, the accuracy of the {\it ab initio} results can still be
improved by semiempirical fitting of the experimental data. This
roughly describes the procedure used in the present work.

It is convenient to present the shift of frequency of an atomic transition
under variation of $\alpha$ in the form
\begin{equation}
	\omega = \omega_0 + Q_1 x,
\label{omega}
\end{equation}
where $x = (\alpha/\alpha_l)^2 - 1, \alpha_l$ is the laboratory value
of the fine structure constant ($\alpha_l = 1/137.036$),
and $\omega_0$ is the experimental value for
frequency at $\alpha = \alpha_l$. Formula (\ref{omega}) is accurate in the
vicinity of $\alpha = \alpha_l$. The purpose of the calculations is to
determine the coefficients $Q_1$. This can be done by small variation of
$\alpha$ in the vicinity of $\alpha_l$:
\begin{equation}
	Q_1 \approx \frac{\omega(\delta x) - \omega(-\delta x)}{2\delta x}.
\label{Q1}
\end{equation}
where $\omega$ are the calculated values of the frequencies.  The lines of
Ni~II observed in quasar absorption spectra correspond to the transitions
between ground state and three states of the $3d^84p$ configuration:
$^2\mbox{F}_{7/2} (E = 57080 \mbox{cm}^{-1})$, $^2\mbox{D}_{5/2} (E =
57420\mbox{cm}^{-1})$, $^2\mbox{F}_{5/2} (E =
58493\mbox{cm}^{-1})$. Energies and $g$-factors of these and other lowest
odd states of Ni~II are presented in Table~I.  One can see from the data
that fine structure multiplets of Ni~II sometimes overlap.  In particular,
the center of the $^2$F doublet lies below the center of the $^2$D doublet.
However, the state $^2$F$_{5/2}$ has higher energy than the
$^2$D$_{5/2}$. This means that if these energies are considered as
functions of $\alpha^2$ there must be a level (pseudo)crossing somewhere
between $\alpha = 0$ and $\alpha = \alpha_l$.  Note that the assignment of
a particular state to a specific fine structure multiplet is best indicated
by the value of their $g$-factors.

Another state of interest, $^2$F$_{7/2}$ is close to the state
$^2$G$_{7/2}$ of a different doublet. Although the values of energies and
$g$-factors of these two states indicate that no level crossing takes place
between $\alpha = 0$ and $\alpha = \alpha_l$, {\it ab initio} calculations
show that such crossing happens in the vicinity of $\alpha = \alpha_l$ (for
$\alpha > \alpha_l$). This level crossing phenomenon makes calculations of
the relativistic energy shifts for Ni~II very difficult. Note that $Q_1$
the coefficients (see eq. (\ref{Q1})) are the slopes of the curve
$E(\alpha^2)$ at $\alpha = \alpha_l$. This slope usually changes sign at
the point of the minimal distance between the levels (the level
(pseudo)crossing point). Therefore, the values of $Q_1$ are very sensitive
to the position of the level crossing. On the other hand, the accuracy of
{\it ab initio} calculations is limited by the incompleteness of the basis
set caused by the large number of valence electrons. Therefore, some
approximations have to be made.  Unfortunately, the positions of the level
crossings and the $Q_1$ coefficients vary significantly if we use different
approximations.  However, the energies and fine structure intervals are
much less affected. In particular, the results of calculations are very
stable for the center energies of the fine structure multiplets. Therefore,
to obtain accurate results for $Q_1$ we have adopted a calculation scheme
which is a combination of the {\it ab initio} calculations with a
semi-empirical fitting. Firstly, we perform the {\it ab initio}
calculations using the Hartree-Fock and configuration interaction
methods. Then, to improve the accuracy, we diagonalize the Hamiltonian
(configuration interaction) matrix for a few close states. The matrix
elements are considered as fitting parameters chosen to fit both the
theoretical energy variation as a function of $\alpha$ in the interval $0 <
\alpha < \alpha_l$ and the experimental energies and $g$-factors at $\alpha
= \alpha_l$.  We consider this scheme in more detail below.

For {\it ab initio} calculations we use the relativistic Hartree-Fock (RHF)
and configuration interaction (CI) methods.  We used a form of the
single-electron wave function which explicitly includes a dependence on the
fine structure constant $\alpha$

\begin{eqnarray}
\psi({\bf r})_{njlm} = \frac{1}{r}
 \left( \begin{array}{c}
f(r)_n \Omega({\bf r}/r)_{jlm} \\ i 
\alpha g(r)_n \tilde{\Omega}({\bf r}/r)_{jlm}
\end{array}
\right).
\label{psi}
\end{eqnarray}
This leads to the following form of the RHF equations
\begin{eqnarray}\label{RHF}
f'_n(r)+\frac{\kappa_n}{r}f_n(r)-[2+\alpha^2(\epsilon_n - \hat V)]g_n(r) = 0\\
g'_n(r)-\frac{\kappa_n}{r}g_n(r)+(\epsilon_n - \hat V)f_n(r) = 0 \nonumber ,
\end{eqnarray}
where $\kappa = (-1)^{l+j+1/2}(j+1/2)$ and $V$ is the Hartree-Fock potential:
\begin{eqnarray}
\hat V f = V_d(r)f(r) - \int V_{exch}(r,r')f(r')dr' .
\label{hfpot}
\end{eqnarray}
The non-relativistic limit can be achieved by reducing the value of
$\alpha$ to $\alpha = 0$.

The ground state configuration of Ni~II is $3d^9$. This is an open-shell
system and the RHF approximation needs to be further specified.  We
presented the contribution of the $3d$ sub-shell to the Hartree-Fock
potential as it was filled ($3d^{10}$) and then subtracted from the direct
part of the potential a spherically symmetric contribution of one
$3d_{5/2}$ electron. The exchange part of the potential remained unchanged.
The single-electron states $4s$, $4p_{1/2}$ and $4p_{3/2}$ are calculated
by removing a contribution of another $3d_{5/2}$ electron from the direct
Hartree-Fock potential.

We carry out CI calculations for 9 external electrons with all core states 
below $3d$ being frozen. In this case the CI Hamiltonian has the form
\begin{eqnarray}\label{CI}
\hat H^{CI} = \sum_{i=1}^9 \hat h_{1i} + \sum_{i<j}^9 \frac{e^2}{r_{ij}}
\end{eqnarray}
where $\hat h_1$ is the one-electron part of the Hamiltonian.

The Hamiltonian (\ref{CI}) does not include important effects of
correlations between the core and valence electrons (see,
e.g. \cite{Kozlov96}). These correlations can be considered as consisting
of two different effects.  One effect is the correlation interaction of a
particular electron with the core electrons (polarization of the core by an
external electron). Another effect is screening of the Coulomb interaction
between the valence electrons by the core electrons. The core polarization
effect affects mostly the single-particle energies (ionization potentials)
of Ni~II. However, the intervals between the excited many-body levels are
not very sensitive to these correlations. Therefore, these correlations are
not so important for the accurate calculations of $Q_1$ and we neglected
them. Screening of the Coulomb interaction affects the interval between
energy levels very strongly. We include the screening in a semiempirical
way by introducing screening factors $f_k$. The factors are introduced in
such a way that all Coulomb integrals of a definite multipolarity $k$ in
the CI calculations are multiplied by the same numerical factors $f_k$.
The values of the $f_k$ are chosen to fit experimental values for the
intervals between states of interest listed in the beginning of this
section.  It turns out that the best fit is achieved at $f_1 = 0.75$, $f_2
= 0.9$ and $f_k = 1$ for all other values of $k$. The results for energy
levels and $g$-factors calculated in this approximation are presented in
Table~I.  Fig. 1 presents the energies of the $^4$F$_{5/2}$, $^2$F$_{5/2}$,
$^2$D$_{5/2}$, $^4$F$_{7/2}$, $^2$G$_{7/2}$ and $^2$F$_{7/2}$ as functions
of $\alpha$. One can see the level (pseudo)crossing at $(\alpha/\alpha_l)^2
= 0.3$ for the $^2$F$_{5/2}$ and $^2$D$_{5/2}$ states and at
$(\alpha/\alpha_l)^2 = 0.9$ for the $^2$G$_{7/2}$ and $^2$F$_{7/2}$
states. Note that the experimental data for the energies and $g$-factors of
the pair of states with $J=9/2$ suggest that there is no level crossing in
the interval $0 < \alpha < \alpha_0$.  This is an indication that we
slightly overestimated the relativistic effects in our {\it ab initio}
calculations.  Therefore, we varied the magnitude of the relativistic
effects to fit the fine structure. The best fit is found for the
relativistic corrections reduced by the factor 0.8. This reduction of the
relativistic effects also gives the correct order of the levels with
$J=9/2$ (no level crossing for $\alpha < \alpha_l$).

As can be seen from Table~I, the calculated fine structure, the intervals
between the levels of the same $J$ and the $g$-factors are reasonably
good. However, the coefficients $Q_1$ are quite sensitive to the position
of the level crossing.  Also, we miss a great part of the correlations
between the valence electrons by restricting our basis set to just five
singe-electron states: $3d_{3/2}$, $3d_{5/2}$, $4s_{1/2}$, $4p_{1/2}$ and
$4p_{3/2}$.  Therefore, to achieve high accuracy in $Q_1$ we should make
one more step. We vary and diagonalize the matrix of the level interaction
to fit all available experimental data for the energy levels and
$g$-factors. Three close states, as presented in Fig.~1, are included into
the diagonalization procedure for both $J=5/2$ and $J=7/2$ states.  It is
convenient to present the interaction matrix in the following form
\begin{equation}
	v_{ij} = e_i \delta_{ij} + q_{ij}\xi (\alpha/\alpha_l)^2.
\label{three}
\end{equation}
Coefficients $q_{ij}$ (at $\xi$=1) are chosen to fit the calculated
behavior of the energies between $\alpha = 0$ and $\alpha = \alpha_l$ as
presented in Fig.~1. Let us remind the reader that the information about
this behavior cannot be extracted from the experimental data and can only
be obtained from {\it ab initio} calculations.
Energies $e_i$ and the scaling factor for the
relativistic effects $\xi$ are chosen to fit the experimental energies and
$g$-factors at $\alpha = \alpha_l$.

It is also important  to estimate 
the uncertainties for the calculated values of the $Q_1$ coefficients.
To start with, we have performed the calculations by fitting only two close
levels (instead of three levels) and compared the results for $Q_1$
 with the three-level calculations. Then we did several fittings by varying the
relative weight factors in the simultaneous fits of the energy levels
 and g-factors. In fact, we minimised the value of
$a\sum (\Delta E/E)^2 + (1-a)\sum (\Delta g/g)^2$ with the different 
 weight factors $a$. Finally, we performed the fitting procedures
 with the different limitations on the values of $e_i$ and $\xi$ 
to keep them close to the results of {\it ab initio}  calculations.
We found that the results for $Q_1$ are reasonably stable and estimated
uncertanties using the spread of these results.

The best fitting parameters together with fitted energies and $g$-factors
are presented in Table 2.
All fitted values are very close to
the experimental results presented in Table 1.

The results for the relativistic energy shifts for the states of interest
are
\begin{eqnarray}
	^2\mbox{F}_{7/2}: \omega = 57080.373(4) - 300(200) x \nonumber \\
	^2\mbox{D}_{5/2}: \omega = 57420.013(4) - 700(200) x \nonumber \\
	^2\mbox{F}_{5/2}: \omega = 58493.071(4) + 800(200) x \nonumber 
\end{eqnarray}
The estimated errors are presented in parentheses, $x =
 (\alpha/\alpha_l)^2 - 1$. The precise values of $\omega_0$ are presented
 in Ref. \cite{Pickering}. These expressions have been used in
 Ref. \cite{Murphy} to search for the variation of $\alpha$.

\section{Conclusion}

It is instructive to compare the relativistic energy shifts for Ni~II with
those of other elements calculated earlier \cite{Dzuba99}. The order of
magnitude of the effect for Ni~II is the same as for its neighbor in the
periodic table, Fe~II \cite{Dzuba99}.  However all energy shifts for Fe~II
are positive and close in value. This is because all the corresponding
transitions are $s - p$ transitions and the values of the relativistic
energy shift is dominated by the contribution of the $s$-electron. The
close values of the relativistic shifts for all frequencies in Fe~II makes
it inefficient to use just these frequencies alone in the search for the
variation of $\alpha$. This is because all possible variation of $\alpha$
will be absorbed by determination of the red-shift parameter $z$.  For this
reason we proposed in Refs. \cite{Dzuba99} and \cite{Dzuba98} to compare
energy shifts in heavy elements, like iron, with the absorption spectrum of
light elements from the same gas cloud. This was first done for the Fe~II
and Mg~II spectra in Ref. \cite{Webb}. The relativistic energy shift in
Mg~II is about ten times smaller than that in Fe~II.  This allowed us to
use the transitions in Mg~II as an ``anchor'' which does not change under
variation of $\alpha$.  Another possibility is to compare absorption
spectra of elements in which the effect is large and opposite in sign,
Fe~II and Cr~II for example \cite{Dzuba98}.  In contrast to Fe~II and other
elements considered in Ref. \cite{Dzuba98}, Ni~II does not need such an
anchor. Since the value of the relativistic shift varies strongly from
state to state -- including change of sign -- both red-shift parameter and
variation of $\alpha$ can be determined by comparing shifts of different
lines of Ni~II alone. This presents a new relatively simple and convenient
way to study possible variation in the fine structure constant in the
absorption spectra of distant quasars.  Consideration of only one element
with shifts of opposite sign should allow one to substantially reduce
systematic errors.


This work was supported by the John Templeton Foundation and the Australian
Research Council.


\begin{table}
\caption{Lowest odd levels of Ni~II (configuration $4d^84p$);
energies, fine structure (cm$^{-1}$) and $g$-factors.}
\label{nickel}
\begin{tabular}{cddddccc}
State & Energy\tablenotemark[1] & 
        Interval\tablenotemark[1] & $g_{exp}$\tablenotemark[1] &
 $g_{nr}$\tablenotemark[2] & Energy\tablenotemark[3] &
 Interval\tablenotemark[3] & $g_{calc}$\tablenotemark[3] \\  
\hline
$^4$D$_{7/2}$ & 51558.1 &         & 1.420 & 1.429 & 58594 &       & 1.4247 \\
$^4$D$_{5/2}$ & 52738.6 & -1180.5 & 1.356 & 1.371 & 59826 & -1232 & 1.3636 \\
$^4$D$_{3/2}$ & 56635.1 &  -896.5 & 1.186 & 1.200 & 60757 &  -923 & 1.1917 \\
$^4$D$_{1/2}$ & 54176.1 &  -541.0 &-0.005 & 0.0   & 61318 &  -561 & 0.0034 \\

$^4$G$_{11/2}$& 53496.8 &         & 1.305 & 1.273 & 60634 &       & 1.2725 \\
$^4$G$_{9/2}$ & 53365.2 &   131.6 & 1.156 & 1.172 & 61009 &  -375 & 1.1892 \\
$^4$G$_{7/2}$ & 54262.7 &  -897.5 & 1.02  & 0.984 & 61823 &  -814 & 1.0153 \\
$^4$G$_{5/2}$ & 55018.8 &  -756.1 & 0.616 & 0.571 & 62542 &  -719 & 0.6049 \\
	
$^4$F$_{9/2}$ & 54557.3 &         & 1.26  & 1.333 & 62228 &       & 1.3042 \\
$^4$F$_{7/2}$ & 55417.9 & -860.6  & 1.184 & 1.238 & 63138 &  -910 & 1.2005 \\
$^4$F$_{5/2}$ & 56075.2 & -657.3  & 0.985 & 1.029 & 63838 &  -700 & 1.0002 \\
$^4$F$_{3/2}$ & 56424.6 & -349.4  & 0.412 & 0.400 & 64259 &  -429 & 0.4153 \\

$^2$G$_{9/2}$ & 55300.0 &         & 1.152 & 1.111 & 63712 &       & 1.1222 \\
$^2$G$_{7/2}$ & 56371.6 & -1071.6 & 0.940 & 0.889 & 65191 & -1479 & 0.9356 \\
			      	   
$^2$F$_{7/2}$ & 57080.3 &         & 1.154 & 1.143 & 65798 &       & 1.1077 \\
$^2$F$_{5/2}$ & 58493.0 & -1412.7 & 0.946 & 0.857 & 67469 & -1671 & 0.9618 \\
			      	   
$^2$D$_{5/2}$ & 57419.7 &         & 1.116 & 1.200 & 66113 &       & 1.1022 \\
$^2$D$_{3/2}$ & 58705.6 & -1285.9 & 0.795 & 0.800 & 67542 & -1429 & 0.8030 \\
\end{tabular}
\tablenotetext[1]{Reference \cite{Moore}}
\tablenotetext[2]{Non-relativistic value for $g$-factors}
\tablenotetext[3]{This work's calculations}
\end{table}


\begin{table}
\caption{Fitting parameters and fitted energies and $g$-factors for the 
states of
most interest of Ni~II. Units for energies and $q_{ij}$ are cm$^{-1}$.}
\label{fit52}
\begin{tabular}{crdcd|lrd}
 $n$ & $e_n$ & $q_{1n}$ & $q_{2n}$ & $q_{3n}$ & State & 
 $E (\alpha = \alpha_l)$ & $g (\alpha = \alpha_l)$ \\
\hline
	\multicolumn{5}{c|}{$J = 5/2, \ \ \xi = 0.6806$} & & \\
 1 & 55678 &  650.41 &  268.63 &  148.21 & $^4$F$_{5/2}$ & 56103 & 1.028 \\
 2 & 57705 &  268.63 &  805.94 &  758.62 & $^2$D$_{5/2}$ & 57382 & 1.111 \\
 3 & 58195 &  148.21 &  758.62 & -746.05 & $^2$F$_{5/2}$ & 58577 & 0.945 \\
	\multicolumn{5}{c|}{$J = 7/2, \ \ \xi = 0.7151$} & & \\
 1 & 55745 & -221.58 &  248.78 &  121.67 & $^4$F$_{7/2}$ & 55513 & 1.118 \\
 2 & 55184 &  248.78 & 1064.33 &  272.29 & $^2$G$_{7/2}$ & 55986 & 0.944 \\
 3 & 58046 &  121.67 &  272.29 & -915.07 & $^2$F$_{7/2}$ & 57424 & 1.138 \\
\end{tabular}
\end{table}
\widetext
\newpage
\input psfig
\psfull
\begin{figure}[b]
\psfig{file=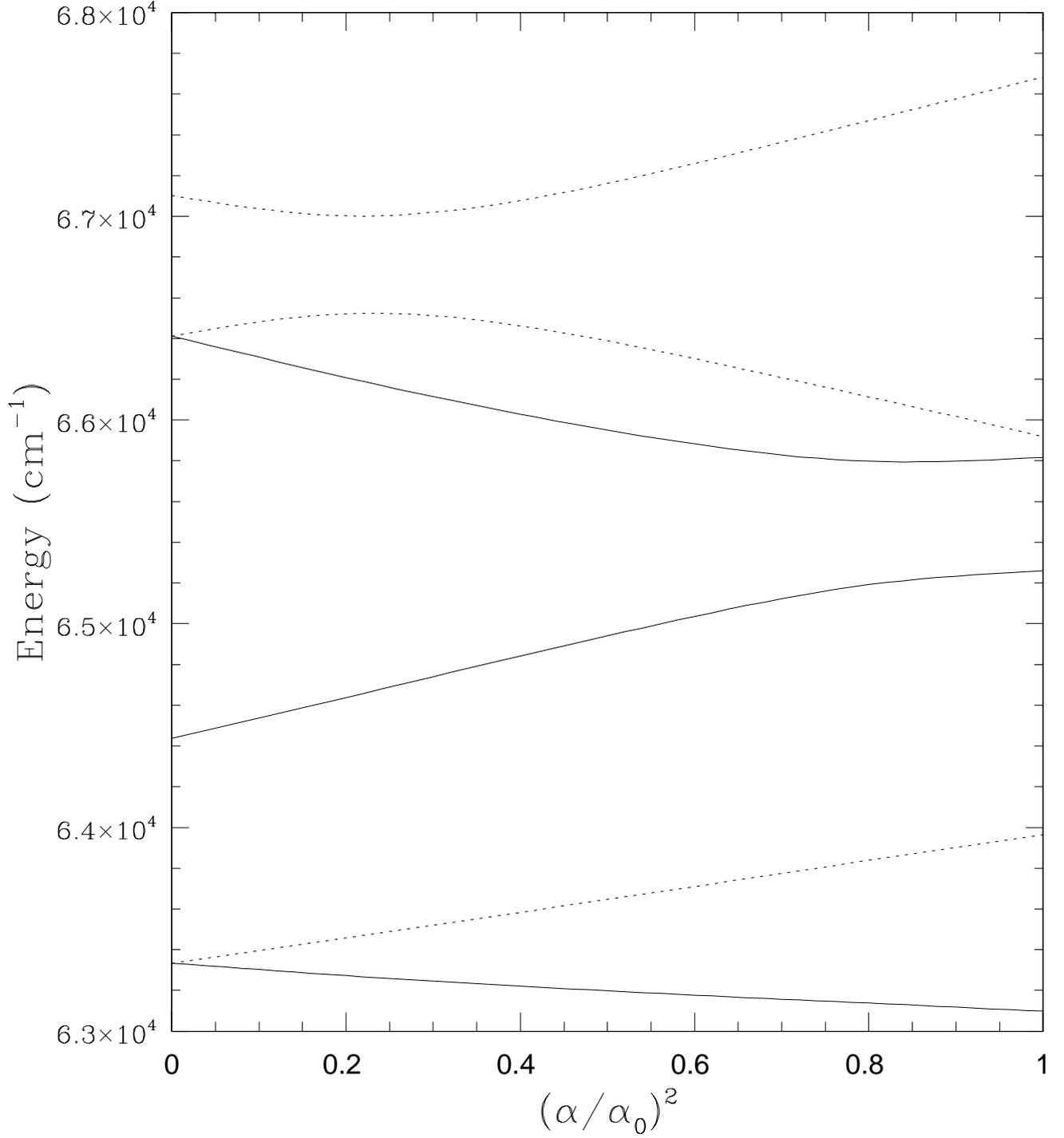, clip=}
\caption{Energy levels of Ni~II with $J=2.5$ (dashed line) and $J=3.5$
(solid line) as functions of $\alpha$. Six states participating in the
semi-empirical matrix diagonalization are shown.}
\label{f1}
\end{figure}
\end{document}